\documentclass[aps,prd,12pt,superscriptaddress,nofootinbib,preprintnumbers,onecolumn,a4paper]{revtex4-1}
\usepackage{amsmath,amssymb}
\usepackage{subfigure}
\usepackage{graphicx}
\usepackage{slashed}
\usepackage{xcolor}
\usepackage[colorlinks,citecolor=blue]{hyperref}

\renewcommand{\emph}{\textit}

\def\mET{\slashed{E}_T}
\begin{document}

\preprint{HRI-RECAPP-2023-009}

 
\title{Search For a Leptoquark and Vector-like Lepton in a Muon Collider}

\author{Nivedita Ghosh}
\email{niveditag@iisc.ac.in (Corresponding Author)}  
\affiliation{Centre for High Energy Physics,
Indian Institute of Science, Bengaluru 560012, India.}

\author{Santosh Kumar Rai}
\email{skrai@hri.res.in}
\affiliation{Regional Centre for Accelerator-based Particle Physics,\\
Harish-Chandra Research Institute,
A CI of Homi Bhabha National Institute,\\
Chhatnag Road, Jhunsi, Prayagraj -- 211019, India.}

\author{Tousik Samui}
\email{tousik.pdf@iiserkol.ac.in}
\affiliation{Department of Physical Sciences,
Indian Institute of Science Education and Research Kolkata,
Mohanpur, 741246, India.}

\begin{abstract}
The proposal for a high-energy muon collider offers many opportunities in the search for physics beyond the Standard Model (BSM). The collider by construction is likely to be more sensitive to the muon-philic models, primarily motivated by the
BSM explanation of muon $(g-2)$ excess and quark flavor anomalies. In
this work, we explore the potential of the proposed muon collider in the context of such models and focus on one such model that extends the Standard Model (SM) with a leptoquark,
a vector-like lepton, and a real scalar. In this model, we propose
searches for TeV scale leptoquarks in $2\mu+2b+\slashed{E}_T$ channel. Notably, the leptoquark can be produced singly at the muon collider with a large cross-section. We have shown that a significant signal in this channel can be
detected at a 3~TeV muon collider even with an integrated luminosity as
low as $\sim 10$~fb$^{-1}$. 
\end{abstract}
\maketitle



\section{Introduction}~\label{s:Intro}
 In the high energy frontier, the highest energies have been achieved in the hadron colliders since the protons could be accelerated to much higher energies than the electrons. The Large Hadron Collider (LHC) has achieved an impressively high beam energy of 6.8 TeV for each colliding proton, making it the highest ever for any collider. 
The available hard scattering energy would be much more in a hadron machine compared to a machine containing electron or positron beams. The hadron machines, however are unable to use 
the full energy of the colliding protons due to their composite nature.
Moreover, while having several advantages in the energy frontier, the hadron collider is plagued by a noisy environment in the form of unwanted hadronic activity and smearing effects from the parton distribution functions (PDFs), which compromises the precision studies.
On the other hand, a much higher energy of the muon beams is achievable through a circular collider due to its significantly smaller synchrotron radiation compared to an electron beam. The muon, being an elementary particle, can, therefore, give high centre-of-mass energies in the hard collisions with a very little energy spread due to the suppressed radiative effects 
of {\it bremsstrahlung} and {\it beamsstrahlung}~\cite{Chen:1993dba,Barklow:2023iav}. This, in turn, helps in the precision measurement of observables and particle properties. Another interesting possibility at these very high-energy muon colliders would be the generation of electroweak gauge bosons as partons of the beam through collinear radiation. These will emerge as electroweak PDFs\,\cite{Han:2020uid,Han:2021kes,Garosi:2023bvq,Frixione:2023gmf}, that can have implications in the study of vector boson fusion (VBF) processes. It is for this reason that the muon collider is also advocated as a ``VBF collider''\,\cite{BuarqueFranzosi:2021wrv}.
 So, to some extent, a muon
collider combines the advantages of $pp$ and $ee$ colliders, {\it i.e.}
the benefits of high energy and high precision\,\cite{Costantini:2020stv,Han:2020uid,Han:2021kes,AlAli:2021let,Accettura:2023ked,MuonCollider:2022xlm}.
Thus, the proposal for a high-energy muon collider by the International Muon
Collider Collaboration (IMCC) is an important development and a recent
growing interest in collider
physics\,\cite{IMCC,Schulte:2019bdl,Delahaye:2019omf,Schulte:2022brl}.

The muon collider will mark a new frontier in collider physics, with high luminosity and beam intensity, including 1 ab$^{-1}$ for a 3 TeV machine and 10 ab$^{-1}$ for a 10 TeV machine\,\cite{Costantini:2020stv,Han:2020uid}. Its uniqueness lies in the muon beams themselves and it will be the first time in human history that a particle collider will be built with a second-generation particle. Therefore, it offers opportunities to directly study muon-related physics\,\cite{Lu:2023ryd,Han:2021lnp,Dermisek:2021mhi,Celada:2023oji,Dermisek:2023rvv,Aime:2022flm,Black:2022cth}.
The potential of physics searches of the proposed muon collider has been explored over the last few years. In most studies, the investigations were primarily aimed at the center of mass energy of 10~TeV or
more~\cite{Delahaye:2019omf}. Furthermore, one expects that the integrated luminosity achievable in this 3~TeV machine will be nearly 1~ab$^{-1}$, comparable to the luminosity achievable at the
14~TeV LHC. Therefore, the early stages of the muon collider could prove crucial in identifying new physics signals that LHC might not be able to probe even with its high luminosity option.

The broad classes of physics studies at the muon collider consist of precision
measurement of electroweak interactions\,\cite{Chakrabarty:2014pja,DiLuzio:2018jwd,Forslund:2023reu},
Higgs properties\,\cite{Han:2020pif,Chiesa:2020awd,Bandyopadhyay:2020otm,Forslund:2022xjq,Han:2021lnp,Reuter:2022zuv,Celada:2023oji,Dermisek:2023rvv},
exploration of new physics sensitivity via higher dimensional effective
operators\,\cite{Casarsa:2021rud}, and new physics searches for well-motivated models beyond the SM
(BSM)\,\cite{Yin:2020afe,Han:2021udl,Huang:2021nkl,Huang:2021biu,Chakrabarty:2022gqi,Ouazghour:2023plc,Belfkir:2023lot,Sun:2023rsb,Maharathy:2023dtp,Guo:2023jkz,Jueid:2023zxx,Chowdhury:2023imd}. 
The `muon-specific' BSM models have an additional advantage in the muon
collider\,\cite{Krnjaic:2019rsv,Jana:2023ogd} as they would lead to direct interactions of the new 
physics sector with the primary colliding beams. The `muon-philic' ($\mu$-philic) BSM models have been 
of wide interest, primarily because of the observed excess in the anomalous magnetic moment of
the muon, namely $(g-2)_\mu$\,\cite{Dermisek:2021mhi}. The latest measurement by the {\sc `Muon G-2'}
collaboration at Fermi National Laboratory (FNAL) combined with the
E989 experiment at the Brookhaven National Laboratory (BNL) stands at
5.1$\sigma$ from its prediction by the Standard Model\footnote{
The 5.1$\sigma$ excess appears in the measured
$(g-2)_\mu$\,\cite{Muong-2:2023cdq,Venanzoni:2023mbe} if we consider the presently available consensus SM prediction given in Ref.~\cite{Aoyama:2020ynm}. However, there are tensions in the Hadronic Vacuum Polarization (HVP)\,\cite{Kurz:2014wya,Davier:2017zfy,Keshavarzi:2018mgv,Colangelo:2018mtw,Hoferichter:2019mqg,Davier:2019can,Keshavarzi:2019abf} contribution to the $(g-2)$ due to the recent lattice QCD based results\,\cite{Blum:2019ugy,Borsanyi:2020mff,Ce:2022kxy,ExtendedTwistedMass:2022jpw,Chao:2022ycy} from BMW collaboration and the $e^+e^-\to \pi^+\pi^-$ data from CMD-3 experiment\,\cite{CMD-3:2023alj}.
}.
In addition, one can also contemplate scenarios where non-trivial physics
may be lurking at electroweak energies. These scenarios are likely to stay hidden because they do not interact with the first generation of the SM fermions which have made up almost all the high energy collider primary beams. The proposed muon collider can be a perfect opportunity to search for such $\mu$-philic models.

In this work, we study a new physics scenario containing a leptoquark, a real scalar, and a pair of vector-like leptons (VLL) in the context of a 3~TeV muon collider. This model was proposed to address various flavor anomalies in the quark sector. Its phenomenological implications
and collider studies have been discussed in Refs.\,\cite{Arnan:2016cpy,Dhargyal:2018bbc,Ghosh:2022vpb,Cheung:2023gwm}. 
The model gives rise to non-universal couplings for different generations of leptons
(quarks) to the VLLs and real scalar (leptoquark). With an appropriate
choice, these non-universal couplings help in explaining the excess in both the
quark flavor violation~\cite{HFAG:2014ebo,Misiak:2015xwa,FermilabLattice:2016ipl,HFLAV:2022esi,LHCb:2022vje} and muon magnetic moment, while the lepton flavor violating (LFV) processes remain within the existing 
experimental bounds. We note that interesting phenomenological studies may exist when the model contains vector-like quarks in addition to VLLs\,\cite{He:1999vp,He:2001fz,Wang:2013jwa,He:2014ora}. We however restrict ourselves to a model containing only VLLs in this work.

We note that there have been recent works on
leptoquark\,\cite{Costantini:2020stv,Asadi:2021gah,Bandyopadhyay:2021pld,Qian:2021ihf,Desai:2023jxh}
and VLL~\cite{Morais:2021ead,Guo:2023jkz} searches in a muon collider. In our study, we perform a search for leptoquarks in the
$2\mu+2b+\mET$ final state, extending our earlier study in the context of LHC in Ref.~\cite{Ghosh:2022vpb}, where we mainly
focused on the sub-TeV masses of the leptoquark, as  LHC is not
very sensitive to heavier masses. However, due to the $\mu$-philic nature of the 
model and the absence of a huge QCD background, a better opportunity to search for such a 
scenario is easier in the muon collider. In
this work, we show that the signature of this model in the same
channel for leptoquark mass of around 2 TeV is within the discovery
range  with just a few fb$^{-1}$ of integrated luminosity at the 
3 TeV muon collider~\footnote{We can safely assume that with the full 1 ab$^{-1}$ integrated luminosity available, the muon collider will clearly outperform the LHC and leptoquarks with mass of upto 2 TeV will be discovered in our model if we search in the $2\mu+2b+\mET$ channel.}. However, as the mass of the leptoquark approaches the
kinematic threshold of the collider, it will be difficult to carry
out the search and require a higher center of mass energy at the muon collider.

The paper is organized as follows. In section~\ref{s:Model}, we
briefly discuss the model we study in this work.
In section~\ref{s:constraint}, we examine how the theoretical and
experimental constraints affect our model parameter space. In section~\ref{s:coll}, we
discuss the possibility of probing the
leptoquark at a 3 TeV muon collider and finally summarize and conclude in
section~\ref{s:conc}.

\section{Model}~\label{s:Model}
The model is an extension of the SM particle content where we add new particles, namely, a real scalar ($S$), a pair of vector-like
leptons ($\ell_{4L}$, $\ell_{4R}$), and a scalar leptoquark ($\Phi$). The only new symmetry introduced 
beyond the SM gauge symmetry is an additional $\mathbb{Z}_2$ symmetry 
with an odd charge to all the new particles. 

Under this new gauge group symmetry of $\mathcal{G}=SU(3)_C \times SU(2)_L \times
U(1)_Y \times \mathbb{Z}_2$, the transformation properties and charges of the new fields
are given in Table~\ref{tab:charges}.
\begin{table}[h]
\begin{tabular}{|c|c|c|c|c|}
\hline
~~Particles~~ & ~~$SU(3)_C$~~ & ~~$SU(2)_L$~~ & ~~$U(1)_Y$~~ & ~~$\mathbb{Z}_2$~~ \\
\hline
$\Phi$ & 3 & 1 & 2/3 & $-1$\\
\hline
$L_{4L}$ & 1 & 2 & $-1/2$ & $-1$\\
\hline 
$L_{4R}$ & 1 & 2 & $-1/2$ & $-1$\\
\hline
$S$ & 1 & 1 & 0 & $-1$\\
\hline
\end{tabular}
\caption{Charges of the new fields under the gauge group
$\mathcal{G}$. All SM particles are even under $\mathbb{Z}_2$.}
\label{tab:charges}
\end{table}

 The gauge invariant Lagrangian density for the new fields and their
interaction with the SM fields is given by
\begin{eqnarray}
\mathcal{L} &\supset& -\mu_\Phi^2 \Phi^\dagger \Phi - \mu_S^2 S^2 -\lambda_{H\Phi} H^\dagger H \Phi^\dagger \Phi - \lambda_{S\Phi} \Phi^\dagger \Phi S^2 - \lambda_{HS} H^\dagger H S^2 - \lambda_\Phi \left(\Phi^\dagger \Phi\right)^2 - \lambda_S S^4 \nonumber\\
            & & -\left\{h_i \bar L_{4R} Q_{Li} \Phi^\dagger + h'_j \bar L_{4R} L_{Lj} S + M_F \bar L_{4L} L_{4R} + h.c.\right\},
            \label{eq:Lag}
\end{eqnarray}
where the SM Higgs doublet, quarks, and leptons are represented by $H$, $Q_{Li}$, and $L_{Lj}$ ($i,j=1,2,3$), respectively. The VLL doublet $L_{4} = (\nu_{4},
 \ell_{4}^-)^T$ consists of the neutral component $\nu_4$ 
and charged component $\ell^-_4$. The couplings $h_i$ and $h'_j$ are responsible for the new interactions between leptoquark-VLL and VLL-real scalar sector. 

The introduction of new neutral scalars generally affects the properties of the SM Higgs boson through mixing 
with these additional scalars. 
However, the mixing will be prohibited due to the unbroken $\mathbb{Z}_2$ symmetry in the new sector, as it prevents all such mixing terms at the tree level. This is due to the fact
that the new real scalar $S$ is forbidden from getting vacuum expectation value (vev) owing 
to its $\mathbb{Z}_2$ odd nature. 
So, the SM Higgs boson becomes massive after the 
spontaneous breaking of the electroweak symmetry. The masses of $\Phi$ and $S$ on the other hand
get additional contributions proportional to the electroweak vev which shifts their masses from $\mu_\Phi$ and $\sqrt{2}\mu_S$ by the
$\lambda_{H\Phi}$ and $\lambda_{HS}$ terms, respectively. The masses then become
\begin{eqnarray}
M_\Phi = \sqrt{\mu^2_\Phi +\frac{\lambda_{H\Phi} v^2}{2}}, \qquad\text{and}\quad M_S = \sqrt{2\mu^2_S + \lambda_{HS}v^2},
\end{eqnarray}
where $v$ is the vev of the SM Higgs doublet.

Furthermore, the couplings of the SM Higgs boson with the other SM particles remain
unchanged at the tree level. Additionally, the new particles are kept
heavier than the SM Higgs boson in order to prevent 
the decay of the Higgs boson to any new modes.

All new contributions to the interaction terms of the SM particles come at
the loop level. Note that the 
leptoquark ($\Phi$) carries a non-zero hypercharge $Y$ as well as color charge and will
contribute to both the $hgg$ and $h\gamma\gamma$ couplings at one loop. This
change in coupling affects the Higgs signal strength~\cite{CMS:2022dwd}.
However, the contributions to the Higgs signal strength in
the gluon-gluon fusion production mode and $\gamma\gamma$ decay channel
are suppressed and are well within the $2\sigma$ limit, provided we keep the
leptoquark mass sufficiently heavy~\cite{Ghosh:2022vpb}.

 The leptoquark and VLL also interact at the tree level with the SM fermions through 
the $h_i$ and $h'_j$ couplings. These two sets of couplings are crucial in addressing the excess observed in the experiments and affect the lepton and quark sector properties through the mediation of the new particles in the loop. These
two couplings $h_i$ and $h'_j$ get
modified after the mixing of quarks and leptons via
Cabibbo-Kobayashi-Maskawa (CKM) and Pontecorvo–Maki–Nakagawa–Sakata
(PMNS) mixing matrices, respectively.  The couplings of the leptoquarks and VLL's with the SM fermions
in the physical (mass) basis become
\begin{eqnarray}
h_i^{\rm ph} = \sum_{m=1}^3 h_m U^d_{mi} \qquad\qquad\text{and}\qquad\qquad h_i^{' \rm ph} = \sum_{n=1}^3 h'_n U^\ell_{ni},
\end{eqnarray}
where $U^d$ and $U^\ell$ are the unitary mixing matrices of down-type
quarks and leptons, respectively. Non-observation of any significant
anomaly in the $K^0$-$\overline{K^0}$ and $B^0$-$\overline{B^0}$
oscillations put restrictions on the couplings of the first two
generations as $h_{1,2}^\text{ph}\simeq 0$~\cite{Dhargyal:2018bbc}.
On the other hand, the observed mass splitting between the physical
states in the $B_s^0$-$\overline{B^0_s}$ sets a constraint
$|h_2^\text{ph} h_3^\text{ph}| \lesssim 0.65$~\cite{Dhargyal:2018bbc,DiLuzio:2017fdq}. The other coupling $h^{\rm ph}_3$ does not have strong upper bound from experiment and we have kept it below perturbativity limit~\cite{Dhargyal:2018bbc}.

One of the central features of this model is that it can account for
the observed excess of the anomalous magnetic moment of muon $(g-2)_\mu$.
This excess is explained by introducing the VLLs $\ell_4$ and the real scalar $S$ at
the one-loop. The lepton flavor violation in $\mu\to e\gamma$ and
$\tau\to\mu\gamma$ can be kept under control by choosing $h'_1$ and
$h'_3$ coupling small. The neutral scalar $S$ can be chosen to be the
lightest one among all the $\mathbb{Z}_2$-odd particles. Thus, the
particle $S$ can act as a dark matter (DM) candidate. The details of
the new physics contribution of this model to $(g-2)_\mu$, implications in flavor physics, and DM aspect of the model have
already been studied and analyzed in~Ref.\cite{Ghosh:2022vpb}, 
and this will be briefly discussed in the next section. 

\section{Theoretical and Experimental constraints}\label{s:constraint}
We implement the model file in SARAH~\cite{Staub:2008uz} and
SPheno~\cite{Porod:2011nf} is used to generate the spectrum files.
We use SSP~\cite{Staub:2011dp} for scanning the parameter space.
For the scanning, we have varied the masses in the following range:
\begin{eqnarray}
M_{\Phi} \in [750:3000]~\text{GeV},\qquad
 M_{\ell_4} \in [102.6:500]~\text{GeV},\qquad M_S \in [65:400]~\text{GeV}
 \label{scanrange}
\end{eqnarray}

The lower bound of 102.6~GeV on the mass of the charged lepton
$M_{\ell_4}$ comes from Large Electron Positron collider
(LEP)\,\cite{ParticleDataGroup:2022pth}. As we will be analyzing the signal at 3 TeV muon collider, we have taken the maximum mass of the LQ to be 3 TeV. The scan range of the VLL is taken to be 500 GeV since it is favored by muon anomaly data. We will see this in the next subsection. As we want our scalar to be a dark matter candidate, for the scan we have taken $M_S < M_{\ell_4}$
and to evade the $h\to$ invisible decay constraints, we have
taken the lower bound on scalar mass to be 65 GeV.

\subsection{Muon anomaly and Lepton Flavor Violation}
In its recent report\,\cite{Muong-2:2023cdq}, the {\sc `Muon G-2'}
collaboration at the Fermilab National Accelerator Laboratory
(FNAL) has announced the experimental measurement of muons
anomalous magnetic moment\,\cite{Aoyama:2020ynm}. This
anomalous part $a_\mu$ is defined in terms of the gyromagnetic
ratio or Land\'e $g$-factor, which is expected to be 2 at the
tree level, as $a_\mu \equiv (g-2)/2$. After taking the effects
of loop corrections, the value of $a_\mu$ in the SM is expected to be more
than zero, and its predicted value, calculated in the SM comes out to be
\,\cite{Aoyama:2020ynm,Aoyama:2012wk,Aoyama:2019ryr,Czarnecki:2002nt,Gnendiger:2013pva,Davier:2017zfy,Keshavarzi:2018mgv,Colangelo:2018mtw,Hoferichter:2019mqg,Davier:2019can,Keshavarzi:2019abf,Kurz:2014wya,Melnikov:2003xd,Masjuan:2017tvw,Colangelo:2017fiz,Hoferichter:2018kwz,Gerardin:2019vio,Bijnens:2019ghy,Colangelo:2019uex,Blum:2019ugy,Colangelo:2014qya}
\begin{equation}
 a^\text{SM}_{\mu} = 116591810(43) \times 10^{-11}. \label{eq:amuSM}
\end{equation}
On the other hand, the experimental uncertainty in the measurements has been brought down by the
{\sc `Muon G-2'} collaboration at
FNAL\,\cite{Muong-2:2015xgu,Muong-2:2021vma,Muong-2:2021ojo} and it has improved significantly to almost half of the uncertainty in the prediction from the SM.
However, the measured central value of $a_\mu$ in its Run-2 plus
Run-3\,\cite{Muong-2:2023cdq} remained almost the same as the
Run-1\,\cite{Muong-2:2021vma,Muong-2:2021ojo}. The new measurement
of $a_\mu$ reads as\,\cite{Muong-2:2023cdq}
\begin{equation}
 a^\text{exp-FNAL}_{\mu} = 116592055(24) \times 10^{-11}.
 \label{Fermi}
\end{equation}
This new measurement from FNAL makes a new combined world average
(combination of FNAL\,\cite{Muong-2:2021ojo,Muong-2:2023cdq} and
older BNL(2006)\,\cite{Muong-2:2006rrc} data)
\,\cite{Muong-2:2023cdq}
\begin{equation}
 a^\text{exp-comb}_{\mu} = 116592059(22) \times 10^{-11}.
\end{equation}

We note here that there are tensions in the SM prediction for the
$a_\mu$ in Eq.~\ref{eq:amuSM} mainly in the
hadronic vacuum polarization (HVP) contribution. In the consensus prediction by Ref.~\cite{Aoyama:2020ynm}, the HVP contributions are calculated using experimental $e^+e^-$
annihilation data\,\cite{Kurz:2014wya,Davier:2017zfy,Keshavarzi:2018mgv,Colangelo:2018mtw,Hoferichter:2019mqg,Davier:2019can,Keshavarzi:2019abf}. An alternative
{\it ab initio} calculation using lattice QCD techniques\,\cite{Blum:2019ugy,Borsanyi:2020mff,Ce:2022kxy,ExtendedTwistedMass:2022jpw,Chao:2022ycy} for
the HVP contribution weakens the tension between the theory prediction and experimental result.
Furthermore, the recent $e^+e^-\to \pi^+\pi^-$ result from the CMD-3 experiment\,\cite{CMD-3:2023alj} disagrees with all previous measurements of this cross-section 
used in the 2020 White Paper\,\cite{Aoyama:2020ynm} and leads to reduced tension with the experimental result.
At present, any firm comparison of the muon $(g-2)$ measurement with the theory is hard to establish and we therefore choose to
work in the paradigm that a 5.1$\sigma$ excess exists,
and a contribution from new physics is needed.

In our model, at one-loop, the new physics (NP) contribution to
$a_\mu$ comes from the scalar $S$ and the VLL and the extra
contribution can be expressed as\,\cite{Arnan:2016cpy,Ghosh:2022vpb}
\begin{eqnarray}
\Delta a_\mu = \frac{m_\mu^2 |h'_2|^2}{8\pi^2 M_{\ell_4}^2}\,f\!\left(\frac{M_S^2}{M_{\ell_4}^2}\right),
\end{eqnarray}
where $m_\mu$ is the mass of muon and loop function 
\begin{eqnarray}
f(x) = \frac{1-6x-6x^2\ln x +3x^2 + 2x^3}{12(1-x)^4}.
\end{eqnarray}

A similar set of Feynman diagrams to muon anomaly will contribute
to the LFV process. Non-observation of any significant
deviation in the charged lepton sector strongly constrains LFV
processes. The strongest bound in the $\mu$--$e$ sector is through
the branching ratio of $\mu\to e\gamma$ process from the MEG
experiment\,\cite{MEG:2016leq}.
Similarly, one also gets constraints from $(\tau \to e\gamma)$ and
$(\tau \to \mu\gamma)$ decay branching ratios (BR). The current
bound on these lepton flavor conversions are~\cite{BaBar:2009hkt} 
\begin{align*}
{\rm BR}(\mu \to e\gamma) < 4.2\times10^{-13}, \ && {\rm BR} (\tau \to e\gamma) < 3.3\times10^{-8} \ ,&&  {\rm BR} (\tau \to \mu\gamma) < 4.4\times10^{-8}. 
\end{align*}

\begin{figure}
\centering
\subfigure[]{\includegraphics[width=8cm, height=6cm]{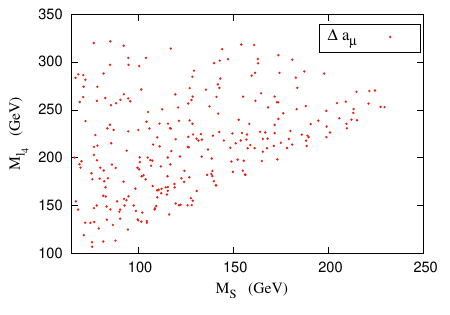}}\hfill
\subfigure[]{\includegraphics[width=8cm, height=6cm]{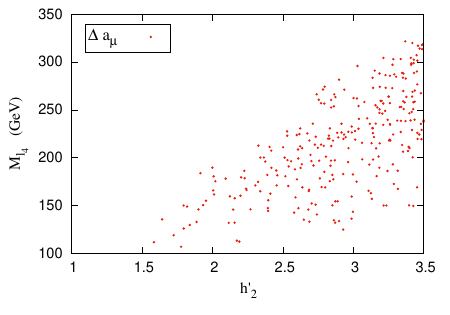}}
\caption{Parameter space allowed by muon anomaly data in
(a) $M_S-M_{\ell_4}$ plane and (b) $h_2'-M_{\ell_4}$ plane. These points also satisfy constraints from LFV measurements. In panel (a), $h_2'$ is varied $\in [1.0,3.5]$, and, in panel (b), the scan takes $M_S \in [65,400]$ GeV. So, different points would possibly have different values of $h_2'$ and $M_S$ in panels (a) and (b), respectively.}
\label{fig:muonanomaly}
\end{figure}

We have plotted the allowed parameter space for muon anomaly and LFV in $M_S$-$M_{\ell_4}$ and $h_2^{'}$-$M_{\ell_4}$ plane in Figs.~\ref{fig:muonanomaly}(a) and \ref{fig:muonanomaly}(b), respectively. From the figure,
we find that the muon anomaly data favors $M_{\ell_4}
\lesssim 330$ GeV  and $h'_2 \gtrsim 1.5$. The constraints coming from the LFV processes can easily be satisfied by tuning the $h'_1$ and $h'_3$ couplings \,\cite{Ghosh:2022vpb}. We have kept the values of $h'_1$ and $h'_3$ couplings ${\cal{O}}(10^{-5})$ and ${\cal{O}}(10^{-2})$, respectively. Another point worth noticing here is that since the leptoquark is odd under $\mathbb{Z}_2$ and has no direct interaction with any SM lepton (see Eq.~\ref{eq:Lag}), it will not contribute to the 
muon anomaly and any LFV processes at one loop. It, however, will have a role in quark 
flavor violation, which puts limits on the leptoquark mass-coupling plane. 

\subsection{Dark Matter}
In our model, by virtue of $\mathbb{Z}_2$ symmetry, the lightest BSM particle can act as a DM. In this work, the scalar $S$ is assumed to be the lightest in the BSM sector. This scalar can, therefore, act as a DM candidate. In the present context, we avoid a detailed discussion of the DM aspect of the model.
In what follows, we will treat the DM as a type of weakly interacting massive particle (WIMP) that was abundant in the early phase of the universe and was in thermal equilibrium with the other SM particles. As the universe cooled down and expanded, the lighter states did not have sufficient thermal energy to produce the heavier DM particles through interactions, and the DM number density became too low to support further interactions and subsequently 'froze out', becoming a persistent relic within the Universe.

The latest measurement of the DM relic density given by Planck~\cite{Planck:2018vyg} is $\Omega_c h^2 = 0.1198\pm0.0012$. The other important measurements come from the DM 
(a) direct detection (DD) and (b) indirect detection experiments.
Non-observation of any significant DM signal in any DD experiments puts an upper limit on the DM-nucleon cross-section. For the DM mass range considered in this work, the strongest constraint comes from XENON-1T~\cite{XENON:2018voc}. On the other hand, the indirect detection experiments constrain the thermally averaged DM annihilation cross-sections $\langle\sigma v\rangle$. For the mass range in our work, the Fermi Large Area Telescope (Fermi-LAT)\,\cite{Fermi-LAT:2013sme,Fermi-LAT:2015att,Fermi-LAT:2016uux} and MAGIC collaboration\,\cite{MAGIC:2011nta,Aleksic:2013xea} provide an upper limit at 95\% C.L. on the DM annihilation cross-section to be $\sim10^{-25}$~cm$^{3}/$s\,\cite{MAGIC:2016xys} in the $\mu^+\mu^-$ channel. 

\begin{figure}
\centering
\includegraphics[width=10cm, height=7cm]{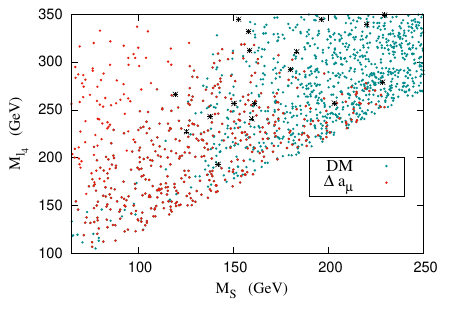}\vspace{-6mm}
\caption{Parameter space allowed by muon anomaly data at 3$\sigma$ (red) and DM relic density (blue and black) in $M_S$-$M_{\ell_4}$ plane. For DM, relic under-abundant points are represented by blue dots, and the black dots satisfy relic density measurement by PLANCK\,\cite{Planck:2018vyg} within 2$\sigma$. For the scan, we keep $M_\Phi \in [750:3000]$ GeV,  $h_2' \in [1.0:3.5]$, $h'_1 =10^{-5}$, $h'_3= 10^{-2}$.}
\label{fig:DM}
\end{figure}
We have performed a scan to satisfy these DM-related measurements, assuming $S$ as the DM candidate. For the scan, we generated the CALCHEP~\cite{Belyaev:2012qa} model file from SARAH~\cite{Staub:2008uz} and pass it through MICROMEGAS~\cite{Belanger:2014vza}, which calculates the DM observables like relic density $\Omega_{\rm DM} h^2$, spin-dependent ($\sigma_{\rm SD}$) and 
spin-independent ($\sigma_{\rm SI}$) cross-sections, and the thermally averaged annihilation 
cross-sections ($\langle \sigma v \rangle)$.
The direct detection constraints are easily satisfied in our model since the scalar $S$ has no direct coupling with the nucleons. On the other hand, in indirect detection, the strongest limits on the parameter are expected from the $\mu^+\mu^-$ channel because of the muon-philic nature of the model. For the parameter range considered in this work, the value of ($\langle \sigma v \rangle$) in this channel lies below the observed value, {\it i.e.} below $10^{-25}$~cm$^3/$s. 
For the relic density, we require that the DM should not be over-abundant in the present universe, {\it i.e.} the relic density should be below the observed value by the PLANCK~\cite{Planck:2018vyg}. In Fig.~\ref{fig:DM}, we show the allowed parameter space in $M_S$ - $M_{\ell_4}$ plane, which satisfies the DM constraints.  In the same plane, the allowed point by the muon anomaly data has also been plotted in Fig.~\ref{fig:DM}. We see that smaller mass differences between the DM and VLL are more favored from DM constraints.

\subsection{EW and collider constraints}
The high value of $h_2'$ needed for the explanation of the observed excess in the $(g-2)_\mu$ affects the $Z\mu^+\mu^-$ coupling the most.
This coupling is measured to be well within SM prediction via electroweak precision observables (EWPO) at the LEP.
The deviation allowed for new physics by the EWPO at 2$\sigma$ is 0.8\%\,\cite{Arnan:2016cpy}, {\it i.e.}
\begin{equation}
\delta g_L^\mu/g_{L,SM}^\mu (M_Z^2) < 0.8\%.
\end{equation}
The loop diagrams contributing to the $g_L^\mu$ coupling are similar to that of the $(g-2)_\mu$ diagrams with the external $\gamma$ being replaced by $Z$ bosons. The NP contribution can be expressed as
\begin{eqnarray}
\frac{\delta g_{L}^\mu}{g_{L,\text{SM}}^\mu}(q^2) = \frac{q^2}{32\pi^2 M_{\ell_4}^2} |h_2'|^2 \,G\left(x \right),
\end{eqnarray} 
where $x =\frac{M_S^2}{M_{\ell_4}^2}$ and $q$ is the momentum transfer, {i.e.} momentum carried by $Z$ boson and 
\begin{eqnarray}
G(x) = \frac{7-36x+45x^2-16x^3+(12x^3-18x^2)\log\,x}{36(x-1)^4}
\end{eqnarray}
For $h'_2=3.0$ and masses $M_S, M_{\ell_4} > 100$~GeV, the changes are less than 0.3\%\,\cite{Ghosh:2022vpb}, which is well within the current limit. We have also checked with $h'_2=3.52$ and have obtained the maximum change to be $\approx$\,0.4\%.

The above discussion on the consistency of the parameter space with experimental measurements indicates that the suitable range for relevant parameters could be
\begin{eqnarray}
h'_{2} \in [1.5,3.5], \qquad  M_\Phi \gtrsim 1000~\text{GeV}, \qquad M_{l_4} \gtrsim M_S \in [150, 250]~\text{GeV}.
  \label{eq:coup}
\end{eqnarray}
With this note, we choose four benchmark points tabulated in Table~\ref{tab:bp} for our collider studies that will be described in the next section.
 \begin{table}
 [!h]
\centering
\begin{tabular}{|c||c|c|c|c||c|}
\hline
& $M_{\Phi}$ (GeV) & $M_{\ell_4}$ (GeV) & $M_S$ (GeV) & \quad$h'_2$\quad\qquad & $\sigma(\mu^+\mu^-\to \mu^+\mu^- b \bar{b} \mET)$ \\ 
\hline

BP1 & 1096.0 & 182.1 & 147.0 & 2.87 &  28.3 fb  \\
\hline
BP2 & 1621.1 & 212.8 & 182.0 & 2.59 &  4.5 fb \\
\hline
BP3 & 1900.2 & 245.4 & 199.6 & 2.88 &  2.2 fb \\
\hline
BP4 & 2367.4 & 258.0 & 207.5 & 2.76 &  0.12 fb \\
\hline
\end{tabular}
 \caption{Benchmark points taken for the collider study and the production cross-section of $\mu^+\mu^- b \bar{b} \mET$ at 3 TeV muon collider. }
  \label{tab:bp}
\end{table}

\section{Collider Searches}~\label{s:coll}
In this section, we discuss the possibility of producing the leptoquark ($\Phi$) and VLL ($\ell_4$) at the muon 
collider and study the signature of the associated production of these $\mathbb{Z}_2$ odd particles. As opposed to the hadron collider, 
the pair production of both the leptoquark and VLL at a 3~TeV muon collider will proceed 
via the photon and $Z$ boson exchange which will have a large $s$-channel suppression. A more promising channel
would be the single production of the leptoquark through the associated production with a VLL and a $b$ quark.
This $2 \to 3$ process is found to generate a larger rate of production cross-section and also
allows a significantly larger range of leptoquark mass that can be probed since the VLL mass is favored to be lighter than 330~GeV to satisfy the muon anomaly excess. Crucially, this 
process forces us to involve all the new particles of the model to participate in the interaction
which would then require all the new model parameters to be included in the analysis.
The VLL decays to a $S$ and a muon and the final decay of the leptoquark gives rise to $2b+2\mu+\slashed{E}_T$:
\begin{eqnarray}
 \mu^+  \mu^- 
 \rightarrow \bar{b} \Phi \ell_4^- 
 \rightarrow \bar{b}\, (b \ell_4^+)\,  (\mu^- S) \rightarrow 
 b \bar{b} \mu^+ \mu^- \mET
 \label{eq:prodphi}
\end{eqnarray}

\begin{figure}
\centering
 \includegraphics[width=10cm, height=7cm]{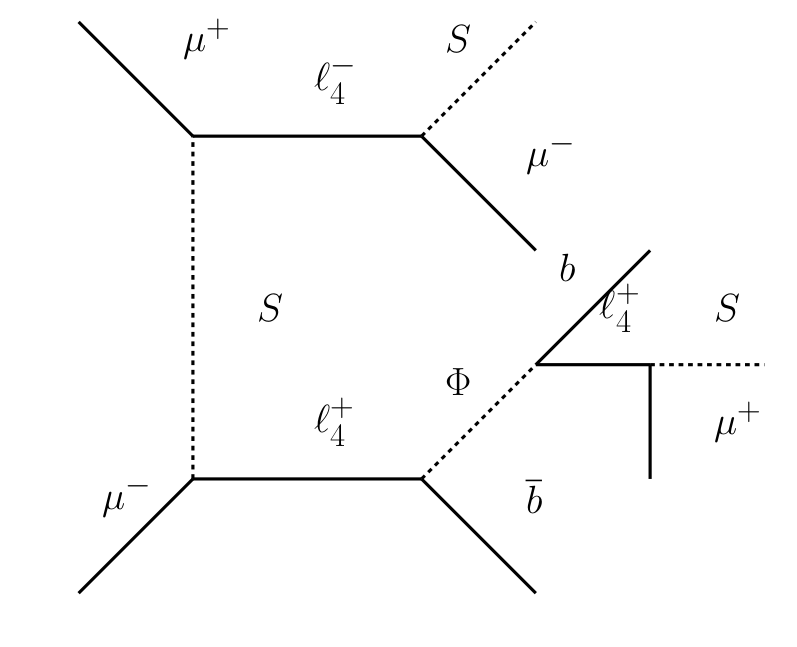}
 \caption{Representative Feynman Diagram for the process $\mu^+\mu^-\to b \bar{b} \mu^+ \mu^- \mET$. }
 \label{fig:feyn}
\end{figure}
A representative Feynman diagram is depicted in Fig.~\ref{fig:feyn}. 
As the leptoquark is produced in association with the VLL and the production of VLL is proportional to the $h_{2}^{'}$ coupling, which is large ($>$\,1.5) to satisfy the muon anomaly, it proves to be an advantage for the production cross-section. The muon-philic nature of this model is reflected in the large production of the above-mentioned signal (Table~\ref{tab:bp}).
 The SM processes that can give rise to similar final states are:
 \begin{itemize}
  \item $\mu^+ \mu^- \rightarrow \mu^+ \mu^- j j \mET$, where jets are misidentified as $b$-jet.
  \item $\mu^+ \mu^- \rightarrow \mu^+ \mu^- b \bar{b} \mET$ which exactly mimics the signal.
 \end{itemize}

 For collider analysis, we have implemented the model file in SARAH~\cite{Staub:2008uz} and have generated the UFO file to generate signal events at MADGRAPH~\cite{Alwall:2014hca}. The spectrum files for the benchmark points are generated using SPheno~\cite{Porod:2011nf}. Both the signal and background events generated in MADGRAPH are passed through PHYTHIA8~\cite{Sjostrand:2014zea} for showering and hadronization. Detector simulation is done in DELPHES-3.5.0~\cite{deFavereau:2013fsa} by editing the default muon collider card accordingly~\cite{ms}. For generating SM backgrounds with hard jets, proper MLM matching~\cite{Hoeche:2005vzu} scheme has been taken into account. We impose the following kinematical acceptance cuts while generating events in MADGRAPH:
 \begin{eqnarray}
   p_T(j,b) &>& 20 ~{\rm GeV}\,; \quad |\eta(j)| < 4.7 \,; \quad |\eta(b)| < 2.5 \ , \nonumber \\ 
 p_T(\ell) &>& 10 ~{\rm GeV}\,, \quad |\eta(\ell)| <2.5 \, , \quad \Delta R_{\ell \ell} > 0.4 \, ,\quad \Delta R_{\ell j} > 0.4 \, , \quad \Delta R_{j j} > 0.4 \,.
  \label{eq:preselect}
 \end{eqnarray}

Tagging $b$-jets in muon collider is not well studied yet as the detector components responsible for measuring impact parameters and displaced vertices are still under research and design. Therefore, we have not used any tagging of the final jets produced at the detector. The jets initiated from the $b$ quarks are treated as normal jets without any $b$-tagging. The cost we have to pay is that we need to consider the other light quark and gluon-initiated jets, which could be reduced greatly with $b$-tagging, in the background.

Furthermore, we note that the traditional search for leptoquark in $2\ell+2j$ channel at the hadron collider\,\cite{D0:2006wsn,ATLAS:2020dsk,ATLAS:2023uox,ATLAS:2023uox,ATLAS:2023kek,CMS:2023qdw} does not constrain our model much\,\cite{Ghosh:2022vpb}. This is because traditional searches expect low missing energy in the final state, whereas our signal consists of high missing energy.
On the other hand, studies prompted by SUSY or VLL searches in the $2\ell+2b+\slashed{E}_T$\,\cite{ATLAS:2016ljb,ATLAS:2016lsr,ATLAS:2016xcm,ATLAS:2017drc,ATLAS:2017mjy} or in the $2\ell+\slashed{E}_T$\,\cite{ATLAS:2017vat,CMS:2018kag,ATLAS:2018ojr} channels can, in principle, constrain our model parameters. Although we did not carry out a detailed scan considering these experimental measurements, we have checked that the four benchmark points are not ruled out by any existing analysis in CHECKMATE~\cite{Drees:2013wra}.
 
\begin{figure}[!h]{\centering
		\subfigure[]{\includegraphics[height = 5.5 cm, width = 8 cm]{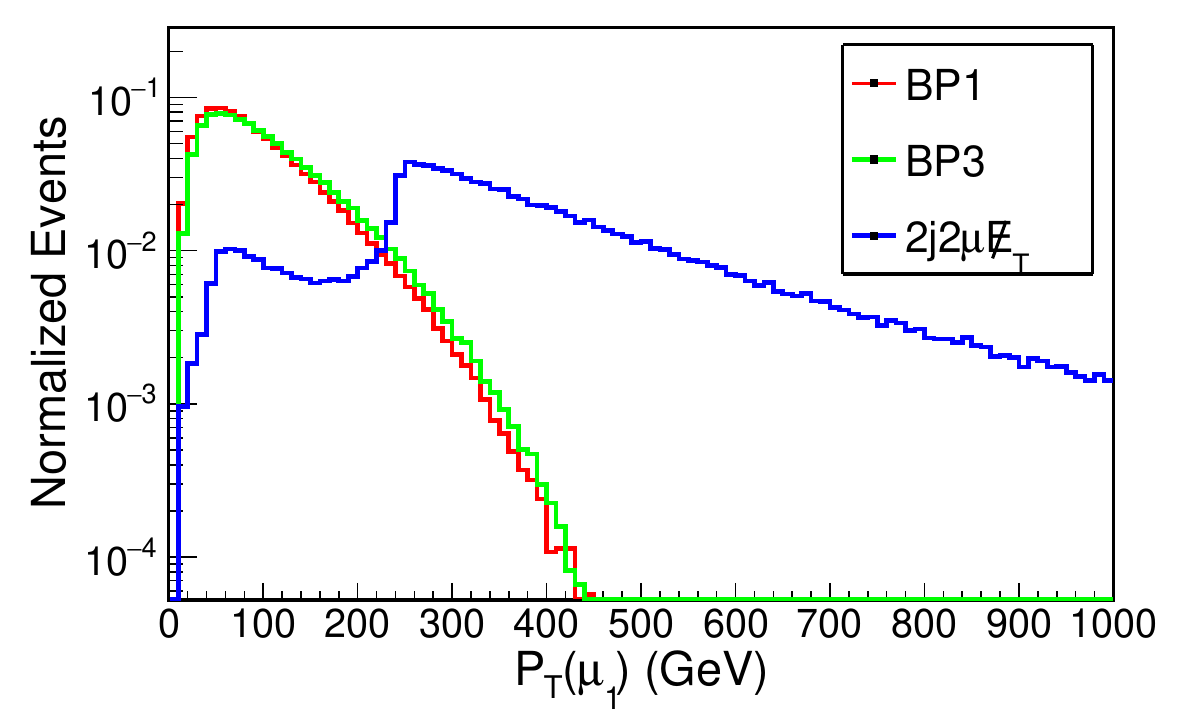}} 
		\subfigure[]{\includegraphics[height = 5.7 cm, width = 8 cm]{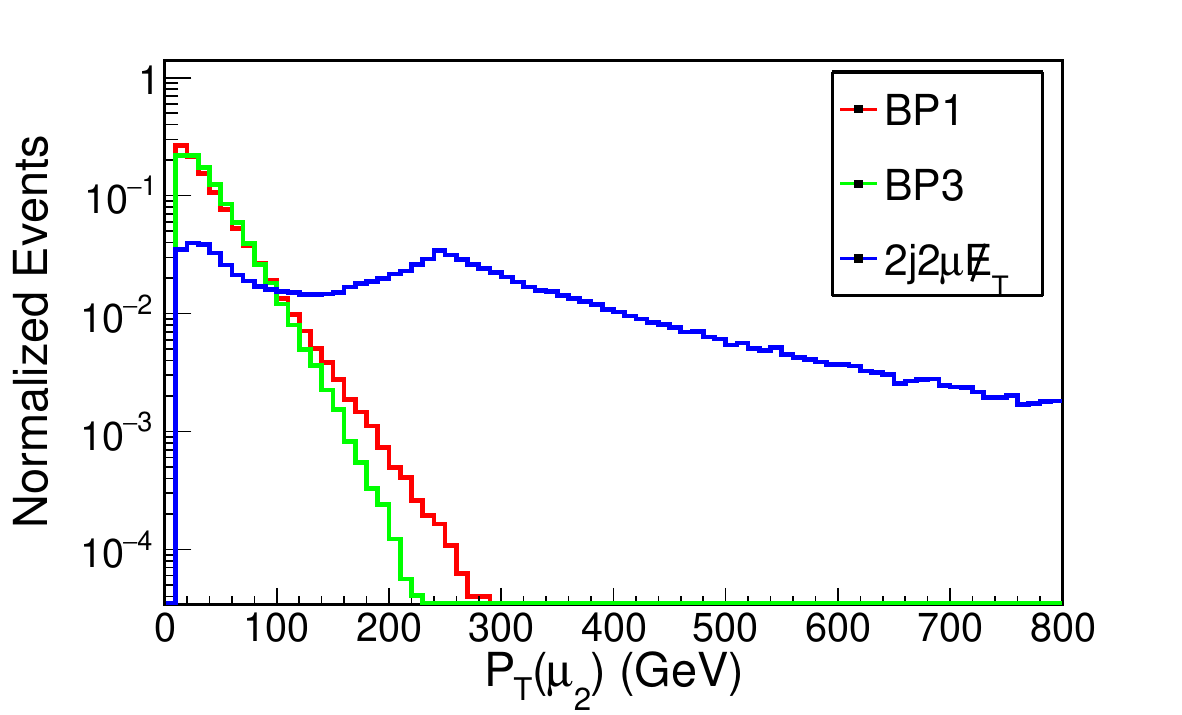}}
		\subfigure[]{
			\includegraphics[height = 5.5 cm, width = 8 cm]{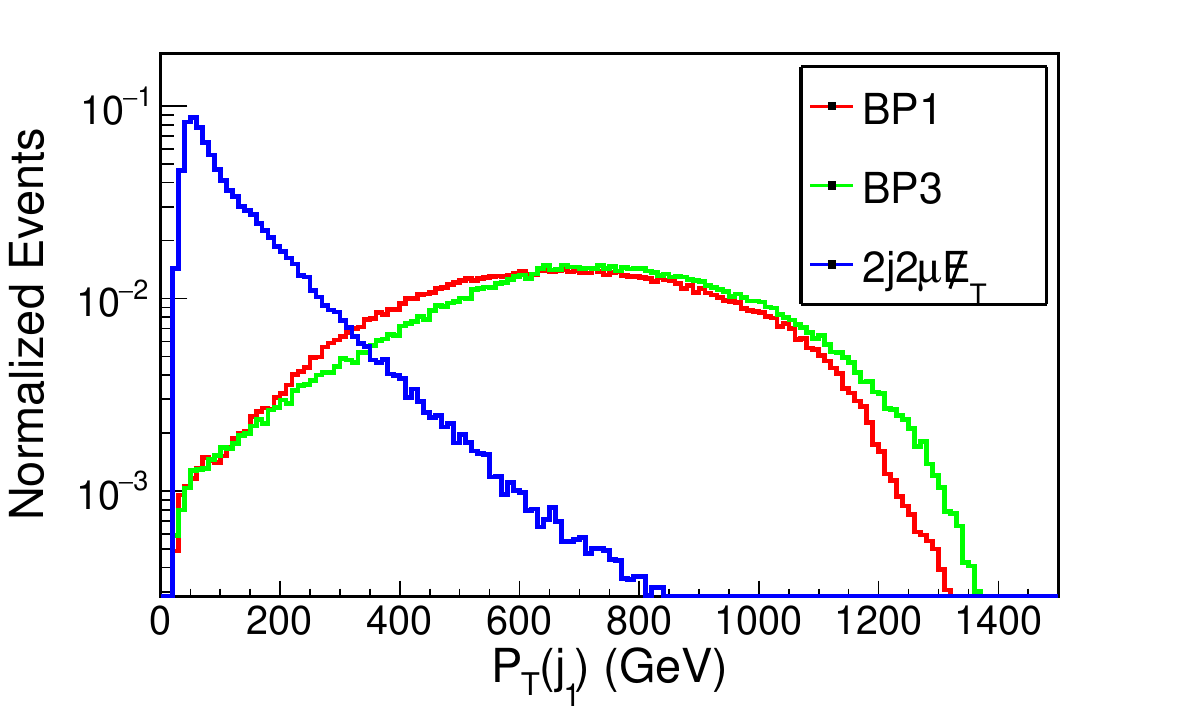}} 
		\subfigure[]{
			\includegraphics[height = 5.5 cm, width = 8 cm]{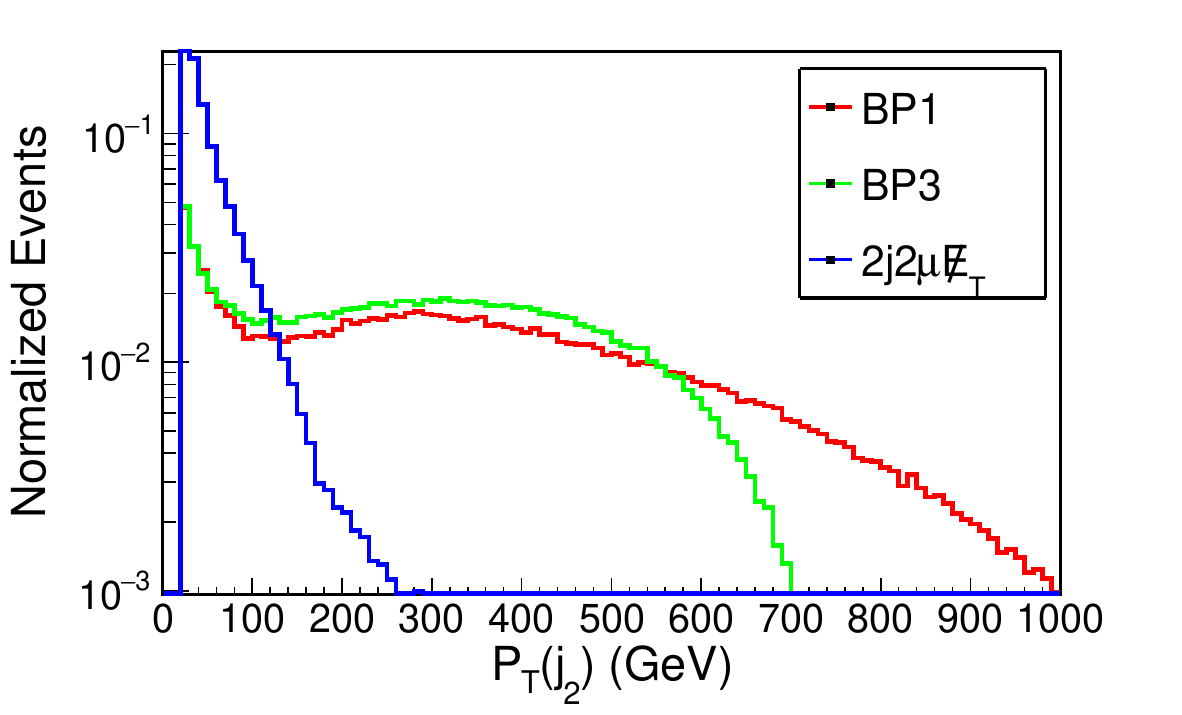}} 
		\subfigure[]{
			\includegraphics[height = 5.5 cm, width = 8 cm]{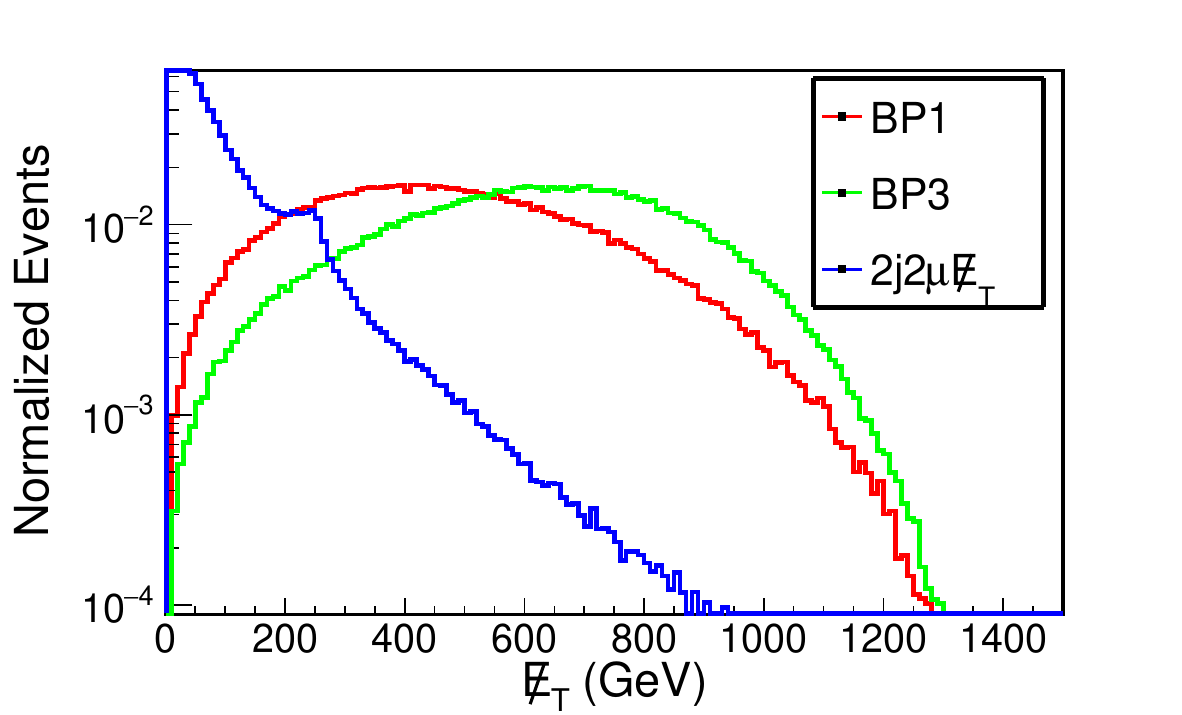}} 
	}
	\caption{Normalized distribution of the kinematic variables
		for the signal and SM backgrounds. }
	\label{fig:kinematics}
\vspace{-4mm}
\end{figure}

 We now provide the details of our cut-based analysis. On top of the preselection cuts discussed in Eq.~\ref{eq:preselect}, we employ the following selection cuts on the kinematic variables:
 
 \begin{itemize}
  \item $p_T(\mu)$: We portray the $p_T$ distribution of the leading and sub-leading muons in Fig.~\ref{fig:kinematics}(a) and Fig.~\ref{fig:kinematics}(b) respectively. To ensure there are exactly two muons, we put a veto on any third muon with $p_T >$ 10 GeV. As we see for backgrounds, the muons tend to populate the higher $p_T$ bins as they come from the hard scattering. We see that $p_T(\mu_1) <$  200 GeV and $p_T(\mu_2) <$  100 GeV helps to reduce the backgrounds.

  \item $p_T(j)$: Momentum distribution for the leading and sub-leading jets are depicted in Fig.~\ref{fig:kinematics}(c) and Fig.~\ref{fig:kinematics}(d). To ensure that the signal contains exactly 2 jets, we reject any third jet with $p_T(j) >$  20 GeV. Compared to the backgrounds, the jets are more boosted for the signal as they come from heavier leptoquark and VLL. Putting cuts of $p_T(j_1) >$ 250 GeV and $p_T(j_2) >$ 150 GeV helps us to reject the backgrounds drastically.
  
  \item $\mET$: For the signal, the $\mET$ comes from the scalar mass S (Fig.~\ref{fig:kinematics}(e)) and hence tends to appear at higher $\mET$ value. We optimize the cut $\mET >$ 100 GeV to enhance the signal over the background.
 \end{itemize}

We summarize the cut flow effects in Table~\ref{tab:cutflow}.
 \begin{table}[htbp!]
	\centering
	\begin{tabular}{|p{3.0cm}|c|c|c|c|p{3.0cm}|}
		\cline{2-5}
		\multicolumn{1}{c|}{}& \multicolumn{4}{|c|}{Number of Events after cuts ($\mathcal{L}=3000$ fb$^{-1}$)} & \multicolumn{1}{c}{} \\ \cline{1-5}
		SM-background  
		& Preselection & $p_T(\mu)$ cut  &  $p_T(j)$ cut    &  $\mET$ cut   & \multicolumn{1}{c}{}
		\\ \cline{1-5} 
          
    $\mu^+\mu^-jj$ & 16376 &  98 & 86  & 38  \\  \cline{1-5}
    $\mu^+\mu^-jj+\mET$ & 5285 & 2296  & 48   &  39  \\ \cline{1-5}
     Total background & 21661 & 2394  & 134  & 77 \\ \cline{1-4} \hline
		
			\multicolumn{1}{|c|}{Signal }  &\multicolumn{4}{|c|}{} &\multicolumn{1}{c|}{Luminosity(fb$^{-1}$) required for 5$\sigma$}  \\ \cline{1-6}
		\multicolumn{1}{|c|}{BP1} & 29877 & 26093  &  19983 & 19390  &\multicolumn{1}{|c|}{0.4}  \\ \hline
		\multicolumn{1}{|c|}{BP2} & 5061 & 4595   &  3447 & 3396  &\multicolumn{1}{|c|}{3.80}  \\ \hline
		\multicolumn{1}{|c|}{BP3} & 2935 & 2567   & 1864  & 1846  &\multicolumn{1}{|c|}{8.7}  \\ \hline
		 \multicolumn{1}{|c|}{BP4} & 146 & 119   &  61 & 61  &\multicolumn{1}{|c|}{1950}  \\ \hline
	\end{tabular}
\caption{The cut-flow for signal and backgrounds for $2\mu
+ 2b + \mET$ channel at the proposed 3 TeV muon collider and the required luminosity to probe with $5\sigma$ significance. }
\label{tab:cutflow}
\end{table}
 We calculate the signal significance by using the
formula\,\cite{Cowan:2010js}
\begin{equation}
\mathcal{S} = \sqrt{2\Big[(S + B) \log\Big(\frac{S + B}{B}\Big)- S\Big]}, \label{eq:signi}
\end{equation}
where $S (B)$ represents the number of signals (background)
events surviving after all the cuts are applied.

We see from Table~\ref{tab:cutflow}, that the three benchmark points with leptoquark mass $<$ 2 TeV can be probed with $5\sigma$ significance with luminosity $<$ 10 $\rm fb^{-1}$. However, once the leptoquark mass approaches the kinetic threshold of the muon collider, the signal significance drops significantly and we need higher luminosity to probe the mass of $\Phi$. We also note that the associated production mode allows the probe of the leptoquark with relatively large masses, whereas the pair production would have restricted the search limits to masses $\lesssim \sqrt{s}/2$. A larger 
center-of-mass energy would further enhance the reach for such leptoquark searches. The DM scalar $S$ also plays a significant role in this search as it mediates the $2 \to 3$ scattering process for leptoquark production. The simultaneous correlation between the flavor sector anomalies and the DM relic abundance also helps in suggesting the regions of parameter space for which this leptoquark production
channel is important. We conclude this section with an optimistic outlook, that the muon collider, if built, will be an excellent opportunity to test 
such $\mu$-philic models of new physics. 
 
\section{Conclusion}~\label{s:conc}
In this work, we have investigated the potential of a muon collider to search for a model with leptoquark, which carries an odd charge under a discrete $\mathbb{Z}_2$ symmetry. The model we consider in this work extends the SM with a $\mathbb{Z}_2$ symmetry along with a set of particles odd under this new discrete symmetry, {\it viz.} a VLL, a real scalar, and a leptoquark. The model offers a new physics explanation for the excess seen in the $(g-2)_\mu$ measurement where the new particles contribute to the loops. In this setup, for a substantial parameter space, we see that the muon anomaly is satisfied when the mass of the VLL $M_{\ell_4} <$ 330 GeV and $h_2' >$ 1.5. The constraint from LFV measurements is satisfied by keeping $h'_1$ and $h'_3$ coupling small. By virtue of the $\mathbb{Z}_2$ symmetry in the BSM sector, the real scalar $S$ can act as a DM, provided the mass difference with VLL is not too large.

Guided by the parameter space, which enables us to explain the experimental excesses as well as provide a DM alternative in the theory, we look at the prospect of
observing the signatures of these new particles at a muon collider. We identify that the associated production of the $\mathbb{Z}_2$ odd leptoquark with the VLL via the exchange of the DM scalar provides an interesting channel to search for these particles.
The proposed search strategy in $2\mu+2b+\slashed{E}_T$ offers a significant signal at a 3 TeV muon collider. Discovery of leptoquarks of mass up to 2~TeV can be easily achieved with an integrated luminosity of around 10~fb$^{-1}$ at the 3 TeV machine. This is a significant improvement compared to the LHC, where only a sub-TeV leptoquark could be searched for in the same channel~\cite{Ghosh:2022vpb}.
Due to the phase-space suppression, a 3~TeV muon collider is not suitable for leptoquark search beyond 2~TeV, for which 10~TeV or higher-energy machines will be more useful.


\section*{Acknowledgement}
The authors acknowledge the support of the Kepler Computing facility maintained by the Department of Physical Sciences, IISER Kolkata, and the RECAPP cluster facility for various computational needs.
SKR would like to acknowledge support from the Department of Atomic Energy, Government of India, for the Regional Centre for Accelerator-based Particle Physics~(RECAPP). NG would like to thank the IISC-IOE fellowship for financial support. NG would like to thank RECAPP, HRI, Prayagraj for hosting while part of the work was going on. TS would like to thank ICTS, Bengaluru and IISc, Bengaluru for their hospitality during the academic visit. 


\providecommand{\href}[2]{#2}\begingroup\raggedright\endgroup


\end{document}